\documentclass[a4paper,twocolumn, nofootinbib,superscriptaddress,aps,prl,,eqsecnum,notitlepage,showkeys]{revtex4-1}
\usepackage{multirow}
\usepackage{graphicx}
\usepackage{bm}
\usepackage{dcolumn}
\usepackage{hyperref}
\usepackage{gensymb}
\usepackage{amsmath}
\usepackage{dcolumn}    
\usepackage{ifpdf}
\usepackage{amssymb,lineno,amsfonts}
\usepackage{graphicx}   
\usepackage{bm}         
\usepackage{bbm}
\usepackage{mathrsfs}
\usepackage{upgreek}
\usepackage{mathtools}
\usepackage{epstopdf}
\usepackage{setspace}
\usepackage{hyperref}
\usepackage{natbib}
\usepackage{esvect}
\usepackage{times}
\usepackage{soul}
\usepackage{amsmath}
\usepackage[usenames,dvipsnames]{xcolor}
\definecolor{med-blue}{RGB}{25,25,112}
\hypersetup{colorlinks, linkcolor={blue},citecolor={Blue}, urlcolor={blue}}
\usepackage [english]{babel}
\usepackage [autostyle, english = american]{csquotes}
\MakeOuterQuote{"}
\begin{document}
\title{Coupled magnetic and ferroelectric states in the distorted honeycomb system Fe$_{4}$Ta$_{2}$O$_{9}$}
\author{Soumendra Nath Panja}
\author{Luminita Harnagea}
\author{Jitender Kumar}
\affiliation{Department of Physics, Indian Institute of Science Education and Research \\ Dr. Homi Bhabha Road, Pune, Maharashtra-411008, India}
\author{P.K.Mukharjee}
\author{R.Nath}
\affiliation{School of Physics, Indian Institute of Science Education and Research, Thiruvananthapuram 695016, India}
\author{A.K.Nigam}
\affiliation{Department of Condensed Matter Physics \& Material Science, Tata Institute of Fundamental Research,  \\Dr. Homi Bhabha Road ,Mumbai 400 005,  India}
\author{Sunil Nair}
\affiliation{Department of Physics, Indian Institute of Science Education and Research \\ Dr. Homi Bhabha Road, Pune, Maharashtra-411008, India}
\affiliation{Centre for Energy Science, Indian Institute of Science Education and Research,
	\\ Dr Homi Bhabha Road, Pune, Maharashtra-411008, India}
\date{\today}
\begin{abstract} 
We report on the magnetic, thermodynamic, dielectric, and pyroelectric  measurements on the hitherto unreported Fe${_4}$Ta${_2}$O${_9}$. This system is seen to exhibit a series of magnetic transitions, many of which are coupled to  the emergence of ferroelectric order, making Fe${_4}$Ta${_2}$O${_9}$ the only genuine multiferroic in its material class. We suggest that the observed properties arise as a consequence of an effective reduction in the dimensionality of the magnetic lattice, with the magnetically active Fe${^{2+}}$ ions preferentially occupying a quasi 2D buckled honeycomb structure. The low temperature $H$-$T$ phase diagram of Fe${_4}$Ta${_2}$O${_9}$ reveals a rich variety of coupled magnetic and ferroelectric phases, in similarity with that observed in the distorted Kagome systems. 
\end{abstract}
\pacs{Pacs}
\maketitle

\section{Introduction} 
Multiferroics- which refer to materials with concomitant magnetic and polar orders in a single system- continues to be at the forefront of contemporary condensed matter physics \cite{Fiebig} . Of particular interest are materials where ferroelectric order arises as a direct consequence of non-trivial spin arrangements, and hence the quest for new and improved multiferroics relies on identifying materials where exotic spin structures (coupled with the anti-symmetric Dzyaloshinskii-Moriya interaction) facilitates the breaking of spatial inversion symmetry \cite{TokuraARCM,Tokura&Seki,CheongReview}   . Geometrically frustrated lattices provide a natural playground for exploring such exotic spin arrangements. This is especially so in low dimensional systems, where a complex interplay between nearest and next-nearest exchange interactions, the single-ion anisotropy, and large spin-orbit coupling conspire to stabilize complex electronic and magnetic ground states. Examples include the S=1/2 spin chain system LiCu${_2}$O${_2}$ \cite{LiCu2O2PRL}, the Kagome-staircase compound Ni${_3}$V${_2}$O${_8}$ \cite{NiVOPRL}, the buckled Kagome system KCu${_3}$As${_2}$O${_7}$(OD)${_3}$ \cite{KCu1} and the triangular lattice system CuFeO${_2}$\cite{CuFeO2PRB}, all of which exhibit novel coupling between the magnetic and polar order parameters. 

The cross-coupling between the magnetic and polar orders is a more generic phenomena, the genesis of which can be traced to Dzyaloshinkii's pioneering work on Cr${_2}$O${_3}$ \cite{Dzyloshinskii}. Here, a linear coupling between these order parameters arises as a direct consequence of the crystallographic symmetry, enabling one to manipulate the electric polarization (magnetization) by an applied magnetic (electric) field\cite{AstrovCr2O3}. In this context, a family of tantalates and niobates of the form $A_{4}X_{2}O_{9}$ (where $A$ = Mn, or Co, and $X$ = Ta or Nb) have recently attracted extensive attention owing to their magnetoelectric properties. First synthesized by Bertaut et.al. \cite{Bertaut}, this family of compounds crystallize in the centrosymmetric Trigonal $P\bar{3}c1$ space group, similar to the prototypical magnetoelectric Cr${_2}$O${_3}$. The structure can be more accurately described to be a variant of the corundum $\alpha$-A$l_{2}$O$_{3}$ with the $A$ and $X$ ions occupying the Al sites in the ratio 2:1. Within this structure, the $A$ site ions occupy two inequivalent crystallographic sites, and its has been suggested that both of them contribute independently to the magnetoelectric effect, with these contributions being of opposite signs in the Co based systems\cite{solovyev}. It is interesting to note that based on structural considerations alone, it was proposed that a member of this extended family Mn${_4}$Ta${_2}$O${_9}$ could harbor a ferroelectric ground state\cite{Abrahams2007}. However, we note that these calculations relied on this system stabilizing in the  $R\bar{3}c$ space group, which is in variance with experimental reports, where all members of the  extended  $A_{4}X_{2}$O$_{9}$ family are reported to be linear magnetoelectrics stabilizing in the trigonal $P\bar{3}c1$ symmetry\cite{Fischer, FangMnTaO, FangNiobate, FangCoTaO,ArimaCoNbO}.

Here, we report on Fe${_4}$Ta${_2}$O${_9}$ - a hitherto unexplored member of this family - using a combination of dc magnetization, specific heat, dielectric and pyroelectric measurements. Unlike all the other members of the  $A_{4}X_{2}O_{9}$ family (which exhibit a solitary para-antiferromagnetic transition), this system exhibits a series of low temperature magnetic transitions. Interestingly, many of these transitions are observed to be associated with the emergence of polar order, making Fe${_4}$Ta${_2}$O${_9}$ the only known multiferroic in this material class. The low temperature $H$-$T$ phase diagram charted out using our measurements shows some similarities with the Kagome staircase systems, making this a novel candidate to explore the intricate coupling between electrical polarization and non trivial magnetic order.  

\section{Experimental}
Polycrystalline samples of Fe$_4$Ta$_2$O$_9$ were prepared using an encapsulation synthesis technique, employing a sealed quartz ampoule which contained a homogeneous mixture of the precursors. Fe granules were used as an  oxygen getter, allowing in situ control of the partial oxygen pressure. A stoichiometric ratio of previously preheated Fe$_2$O$_3$ and Ta$_2$O$_5$ was manually ground for a few hours in a glove box filled with argon, following which the mixture was transferred into a ball mill container and further ground under a static argon atmosphere. A preheated quartz ampoule ($\approx$ 25 $cm^3$) with two alumina crucibles was used for the sintering treatment. The larger alumina crucible was used to accommodate the oxygen getter (Fe grains, 1-2mm), and the smaller one (filled with the mechanically homogenized mixture of precursors) was placed inside the larger crucible. 
\begin{figure}
	\centering
	\hspace{0.15cm}
	\includegraphics[scale=0.40]{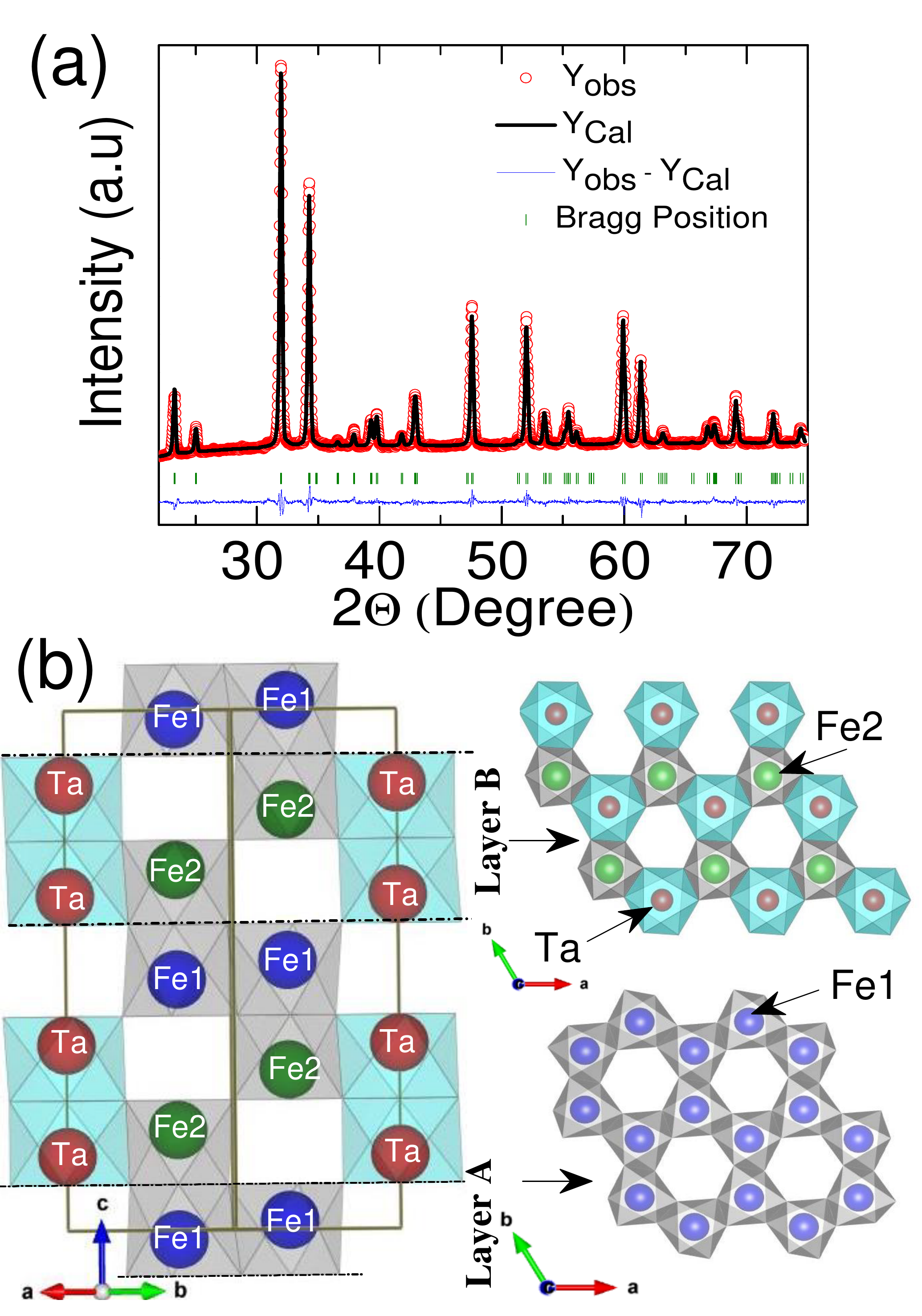}
	\caption{ (Color online ) (a) The Rietveld refinement of room temperature x-ray diffraction data of Fe${_4}$Ta${_2}$O${_9}$. This corresponds to a fit with $R$ parameters of
$R{_{wp}}$=9.13, $R{_{e}}$=5.55, and $\chi{^2}$=2.7. (b) depicts the structure of this system, which can be viewed as a stack of alternating $A$ and $B$ layers along the crystallographic $c$ axis.}
	\label{Fig1}
\end{figure}

The ampoule charged in this fashion was evacuated, sealed under vacuum, and slowly heated to 1100$^o$C and kept there for 48 hours. The oxygen released during the reduction of Fe$^{3+}$ to Fe$^{2+}$ is continuously absorbed by the getter, thus stabilizing a pure Fe$_4$Ta$_2$O$_9$ end product avoiding parasitic phases like FeTa$_2$O$_6$, FeTaO$_4$ and Fe$_3$Ta$_2$O$_8$.  Phase purity of the compound was determined by room temperature X-ray powder diffraction using Cu $K_\alpha$ radiation. Low temperature powder XRD measurements were carried out down to 15 K using a low-$T$ attachment (Oxford Phenix) to the diffractometer (PANalytical). The powder XRD pattern was analyzed by the Rietveld method using the Fullprof suite \cite{Fullprof}. Elemental composition and homogeneity was further confirmed using an energy dispersive X-Ray spectrometer (Ziess Ultra Plus). Specific heat and  DC magnetization were measured using a Quantum design PPMS and a MPMS-XL SQUID magnetometer respectively. Dielectric measurements were performed by using an Alpha-A high performance frequency analyzer from Novocontrol Technologies. The pyroelectric measurements were performed in the parallel plate geometry using a Keithley sourcemeter (Model 2612B) and a Picoammeter (Model 6482). Zero field pyroelectric measurements were performed in a Closed Cycle Refrigerator, and the magnetic field dependent dielectric and pyroelectric measurements were performed using the Manual Insertion Utility Probe of the MPMS-XL magnetometer.  The  polarization was deduced from the integration of measured pyroelectric current over time according to the formula $P =\int  (I/A) dt$ , where $A$ is the area of  sample, $I$ is the measured pyrocurrent and $t$ is the time. 

 \section{Result and Discussions}
\begin{figure}
	\centering
	\includegraphics[scale=0.245]{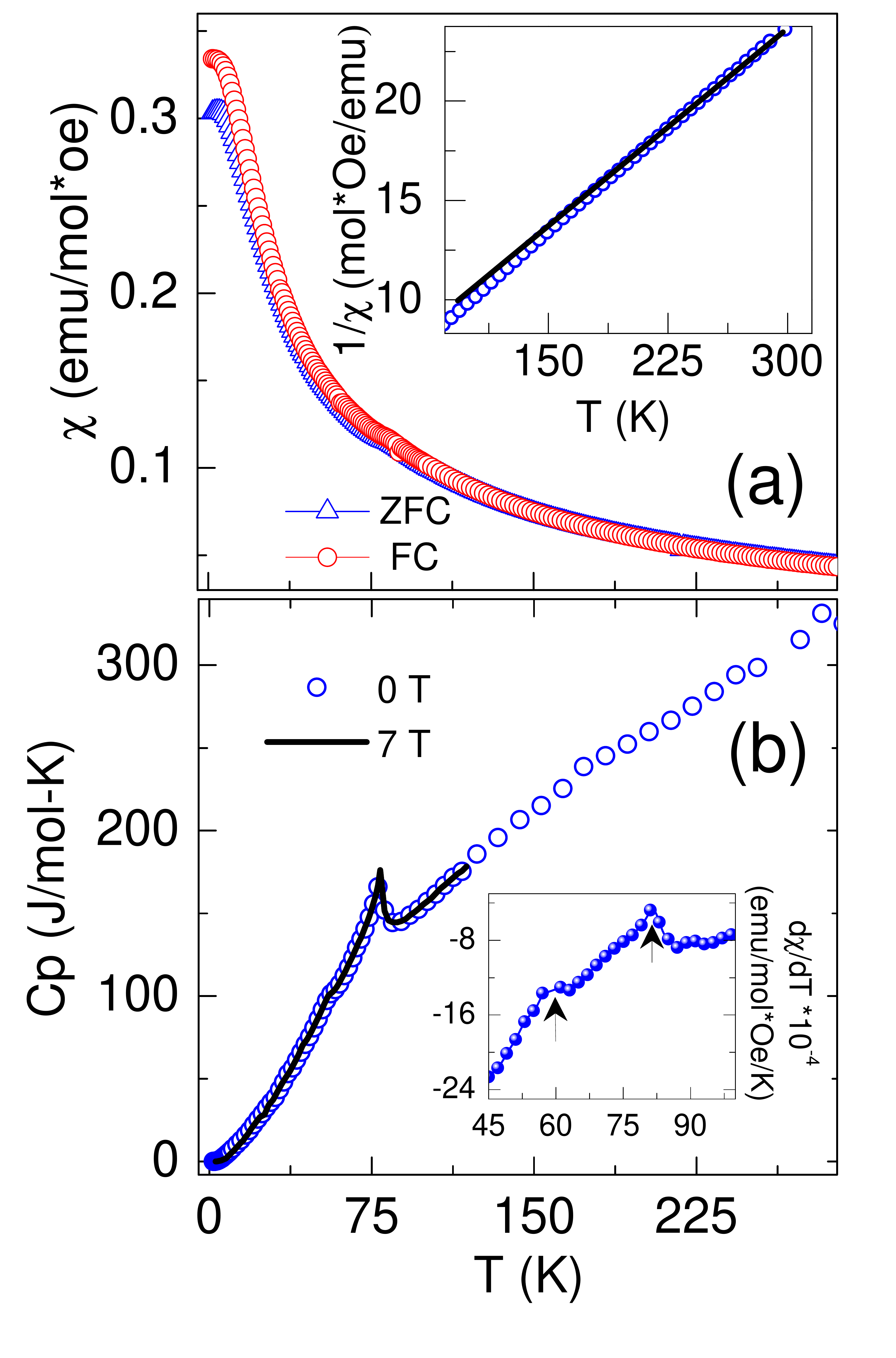}
	\caption{(color online) (a) Temperature dependence of the DC magnetic susceptibility as measured in ZFC and FC protocols at an applied magnetic field of 100 Oe. The inset shows a Curie-Weiss fit to this data in the high temperature regime. The inset of (b) depicts $d\chi/dT$ vs $T$ clearly indicating the presence of transitions at $T_1$ (80K) and $T_2$ (60K). The main panel of (b) depicts the temperature dependence of  Specific heat $C_{P}$ as measured at 0 and 7 Tesla. }
	\label{Fig2}
\end{figure}

 \begin{table}
 	\caption{ Structural Parameters of Fe${_4}$Ta${_2}$O$_{9}$ as determined from the Rietveld analysis of room temperature X-ray diffraction data. }
\begin{tabular}{c c c c c }
		\hline
		\multicolumn{5}{c} {Fe${_4}$Ta${_2}$O$_{9}$ } \\
		\multicolumn{5}{c}{Temperature = 296 K} \\
		\multicolumn{5}{c}{Space Group : $P\bar{3}c1$ (No. 165 )} \\
		\multicolumn{5}{c}{Crystal system: Trigonal }\\
		\multicolumn{5}{c}{ a= 5.232(4) ${\AA}$}\\
		\multicolumn{5}{c}{ b=5.232(4) ${\AA}$ }\\
		\multicolumn{5}{c}{c= 14.238(1) ${\AA}$ }\\
		\multicolumn{5}{c}{${\alpha}$ = ${\beta}$ = 90 \degree ,${\gamma}$ =  120 \degree}\\
		\hline
		\hline
		Atom & Wyckoff & x/a &y/b & z/c \\
		\hline
		Fe1  & 4d & 0.3333 & 0.6667  & 0.5172 \\
		Fe2  & 4d & 0.3333  & 0.6667  & 0.3079\\
		Ta  & 4c & 0 & 0 & 0.8567 \\
		O1  & 6f & 0.3112 & 0 & 0.2500 \\
		O2  & 12g & 0.3288 & 0.2869 & 0.0817 \\
		\hline
	\end{tabular}
	
	\label{Table1}
\end{table}
The Rietveld refinement of the room temperature powder X-Ray diffraction data is shown in Fig. \ref{Fig1}(a). In similarity to the other members of the $A_{4}$Ta$_{2}$O$_{9}$ family, this system is also seen to crystallize in the centrosymmetric Trigonal ($P\bar{3}c1$) symmetry. Along the crystallographic $c$ axis, the structure of Fe${_4}$Ta${_2}$O$_{9}$ can be visualized to be made up of two distinct layers. The first layer ($A$) comprises of hexagonal rings of edge sharing Fe1O${_6}$ octahedra, with the adjacent layer ($B$) being made up of alternating edge sharing Fe2O${_6}$ and TaO${_6}$ octahedra, as is schematically shown in Fig.\ref{Fig1}(b) \cite{Castellanos}.  The structural details as obtained from the Rietveld refinement of room temperature diffraction data are summarized in Table\ref{Table1}. 
The temperature dependence of DC magnetization as measured at 100 Oe  in the Zero field cooled (ZFC) and Field cooled (FC) protocols is shown in Fig.\ref{Fig2}(a).  On cooling from room temperatures, two magnetic transitions at 80 K ($T{_1}$) and 60 K ($T{_2}$) are observed. These are more clearly evident from the temperature dependence of $d\chi/dt(T)$ as is shown in the inset of Fig.\ref{Fig2} (b).  On further cooling, a splitting between the ZFC and FC curves is observed, with the magnetization increasing monotonically till a low temperature magnetic transition ($T{_3}$) at 5K. The inverse of the dc magnetic susceptibility [$\chi{_{dc}}{^{-1}}(T)$] is linear only above 150 K, as is shown in the inset of Fig.\ref{Fig2}(a), indicating the presence of short range correlations well above the transition temperature. The linear fit to $\chi{_{dc}}{^{-1}}(T)$ gives a Curie-Weiss temperature $\theta{_{CW}}$ of -45.33K, indicating mixed ferromagnetic and antiferromagnetic interactions. Surprisingly, the calculated $\mu{_B}$/Fe$^{2+}$ as deduced from the Curie-Weiss fit is only 2.69 $\mu_{B}$, which is significantly less than the spin only ($S$=2) value of 4.89 $\mu{_B}$ expected from a Fe${^{2+}}$ ion in its high spin state.  This is also in variance with that observed in other members of the $A_{4}$Ta$_{2}$O$_{9}$ family, where the effective $\mu{_B}$/magnetic ion is always seen to be larger than the corresponding spin only value. This has been ascribed to indicate the presence of a unquenched orbital momentum contribution, which is also intimately related to the observed linear magnetoelectricity in those systems \cite{Spladin,solovyev,ArimaCoNbO}. Our observations in Fe$_{4}$Ta$_2$O$_9$ indicates that Fe$^{2+}$ coexists in both the high spin ($S$=2) and the low spin ($S$=0) state  in this system, and our Curie Weiss fit indicates that the ratio of the high spin:low spin Fe$^{2+}$ species is in the ratio $\approx$ 0.55:0.45. Specific heat measurements reconfirm the presence of transitions at 80 K and 60 K, as is shown in Fig.\ref{Fig2} (b), and  measurements performed in the presence of a magnetic field also indicates that both of these anomalies are relatively insensitive to the application of magnetic fields up to 7 Tesla. 
\begin{figure} 
	\centering
	\includegraphics[scale=0.50]{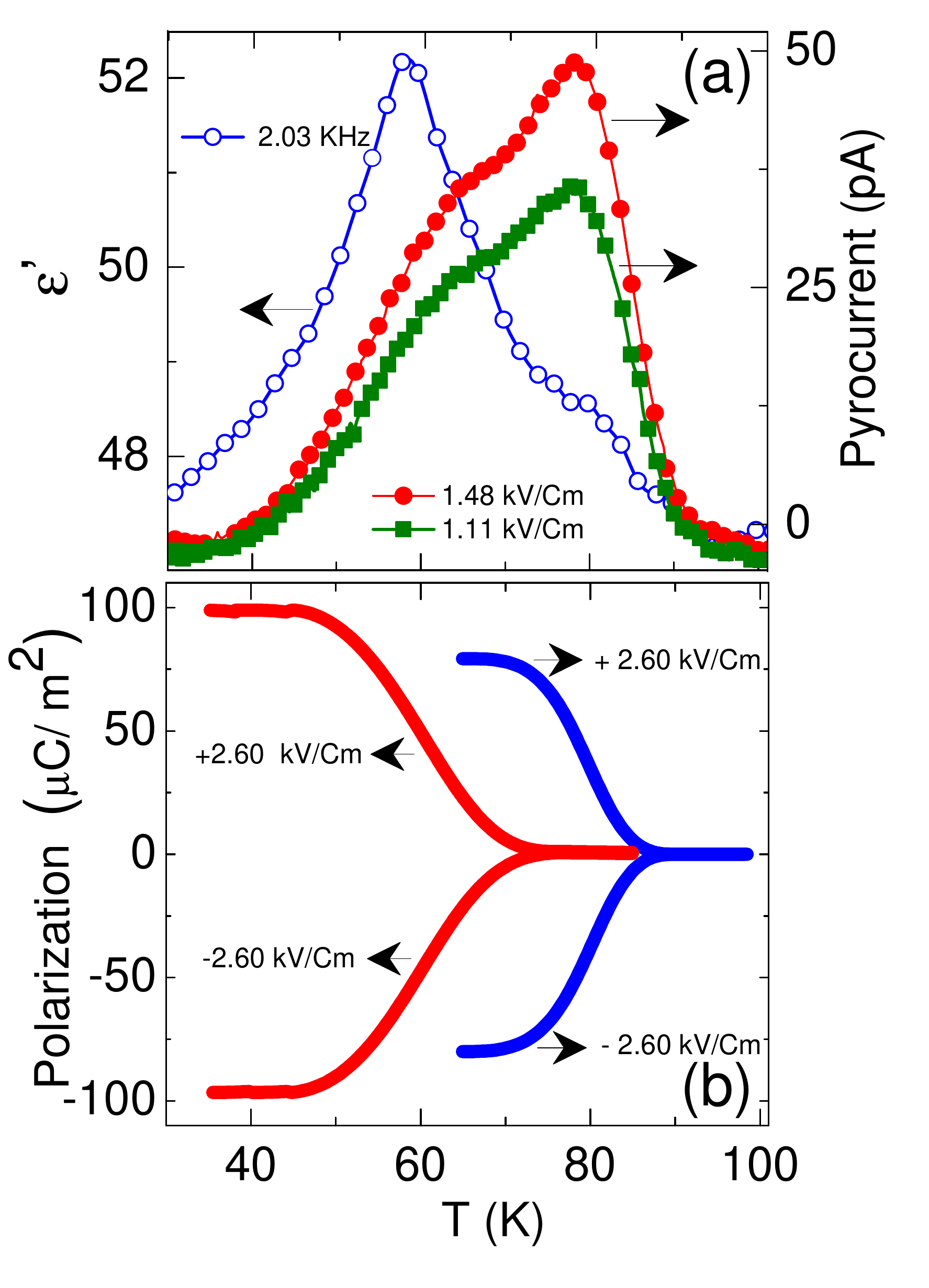}
	\caption{ (Color online )(a) depicts the dielectric constant ($\epsilon'$) and the pyrocurrent in the vicinity of the transitions at $T_1$ (80K) and $T_2$ (60K). (b) depicts the flipping of the polarization associated with these transitions on reversal of the poling electric field, indicating that they correspond to robust independent ferroelectric states.} 
	\label{Fig3}
\end{figure}

Dielectric measurements performed under zero magnetic field exhibits a weak anomaly at $T{_1}$=80 K, and a pronounced  feature at $T{_2}$=60 K [Fig.\ref{Fig3}(a)]. Interestingly, pyrocurrent measurements also exhibit the appearance of features at both $T{_1}$ and $T{_2}$, indicating that both these transitions could be accompanied by ferroelectric order. This is further confirmed by measurements where the pyrocurrent was independently evaluated across each of these transitions, as shown in Fig. \ref{Fig3} (b). This involved poling the sample from 70K (ie $T{_2}<T<T{_1}$) revealing a single peak in the pyrocurrent at 60K, and subsequently poling the sample from 85K to 65K, revealing a peak in the pyrocurrent at 80K. Moreover, the sign of the pyrocurrent could be reversed in both the cases by reversing the direction of the poling field, indicating the presence of two independent robust ferroelectric states at $T{_1}$ and $T{_2}$.  Our low temperature X-Ray diffraction measurement indicate that the crystallographic symmetry of Fe${_4}$Ta${_2}$O$_{9}$ remains invariant down to 15 K. Thus, the two anomalies observed in the magnetization and specific heat measurements can be ascribed to arise from magnetic ordering alone. The concomitant presence of magnetic and polar orders as evidenced by our magnetic and pyroelectric measurements indicate  that Fe${_4}$Ta${_2}$O$_{9}$ is a genuine multiferroic. We note that this is in contrast to that observed in other members of the $A_{4}$Ta$_{2}$O$_{9}$ family, all of which are reported to exhibit a solitary para-antiferro magnetic transition \cite{FangNiobate} and are reported to be linear magnetoelectrics - with polar order only being observed in the presence of an applied magnetic field. 
\begin{figure}[b]
	\centering

	\includegraphics[scale=0.48]{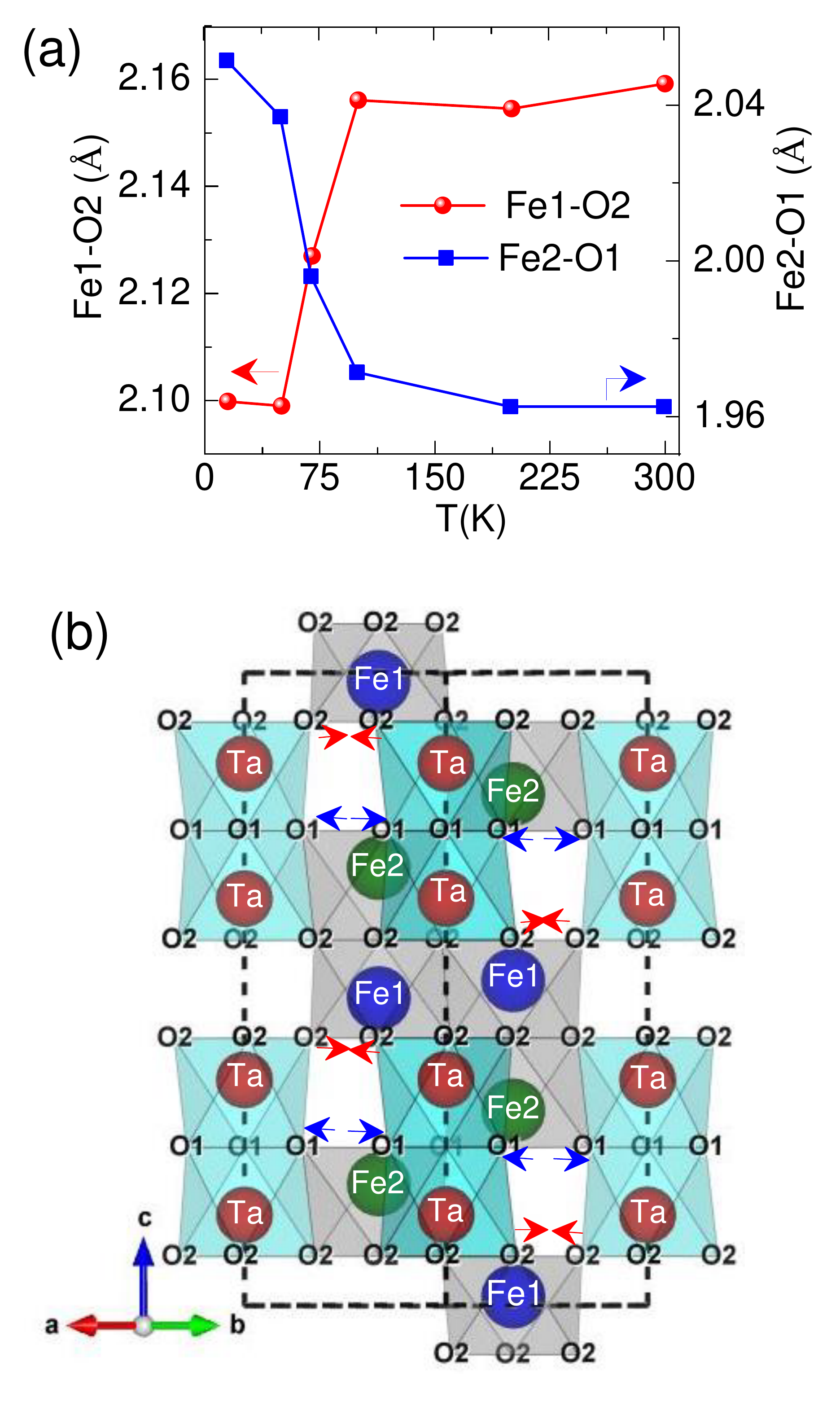}
	\caption{ (Color online ) (a) depicts the variation of the Fe1-O and Fe2-O bond lengths across the magnetic transition. (b) shows the crystallographic structure of Fe${_4}$Ta${_2}$O${_9}$ below the magnetic transition, where the buckling of the FeO polyhedra is observed.}
	\label{Fig4}
\end{figure}
\begin{figure}
	\centering
	\includegraphics[scale=0.43]{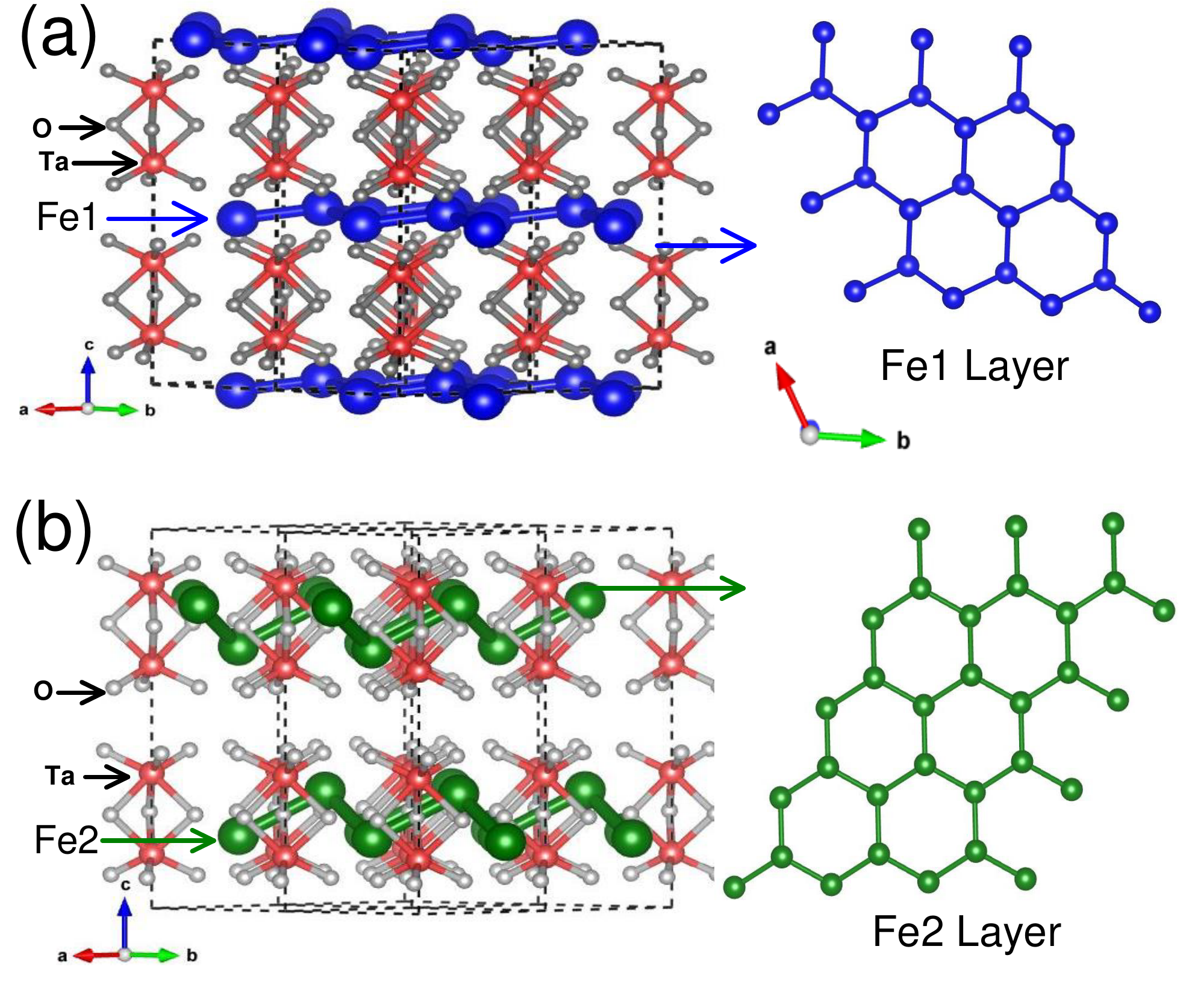}
	\caption{(color online) a) depicts the buckled honeycomb lattice formed by the Fe1 ions separated by the non magnetic TaO${_6}$ octahedra, (b) depicts the  zig-zag chain of the non-magnetic  Fe2 ions. Considering the fact that the Fe1-O bond length is larger than the Fe2-O one, we speculate that the magnetic (S=2) Fe${^{2+}}$ ions preferentially occupy the buckled honeycomb lattice, and the non-magnetic (S=0)  Fe${^{2+}}$ ions preferentially occupy the zig-zag chain. }
	\label{Fig5}
\end{figure}
 
\begin{figure}
	\centering
	\includegraphics[scale=0.44]{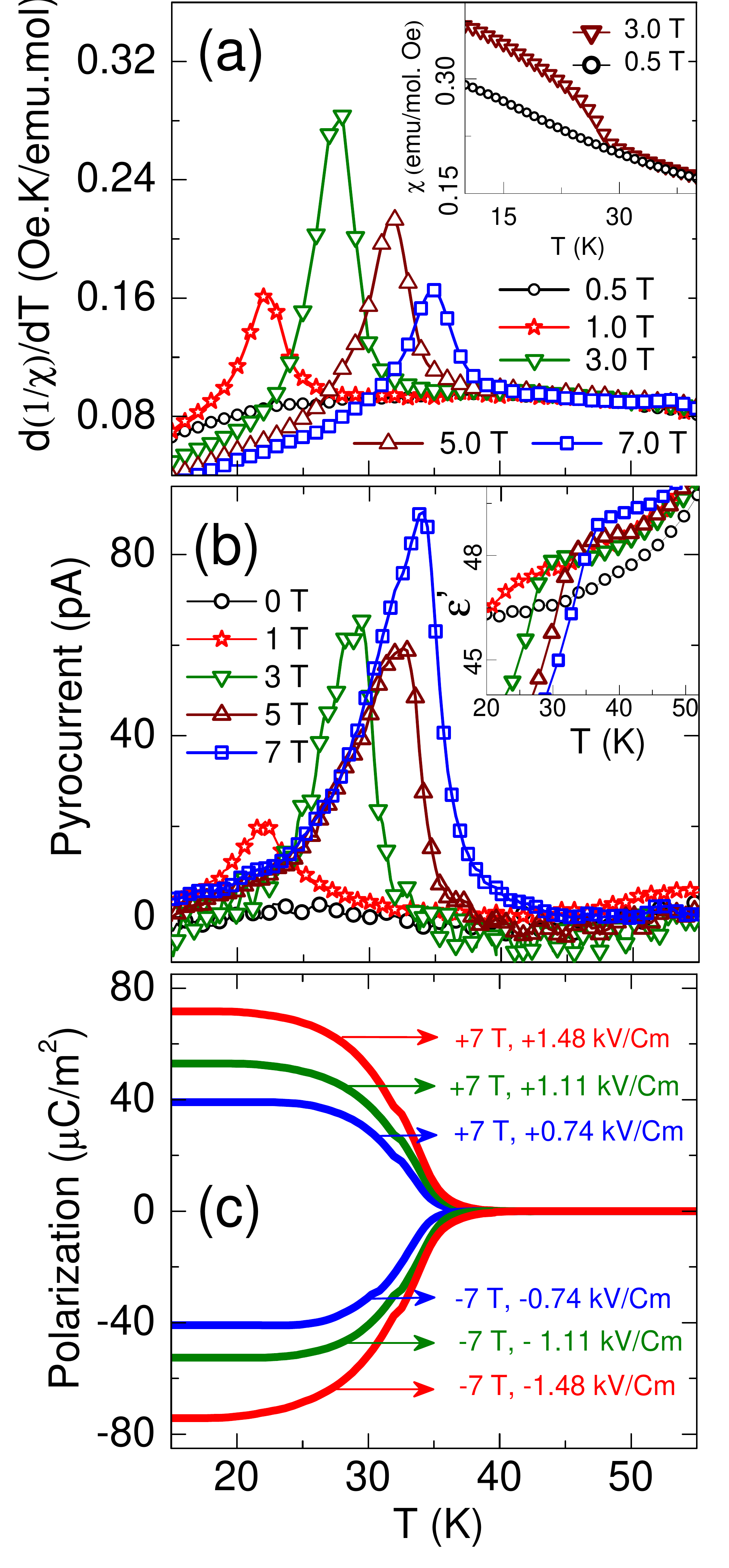}
	\caption{ (color online) Inset of (a) depicts the DC magnetic susceptibility as measured at 5kOe and 3 Tesla, indicating the presence of a magnetic field induced transition. This is seen more clearly in the plots of the derivative (d(1/${\chi}$))/dt) vs $T$ as shown in the main panel. Inset of (b) shows the temperature dependence of the dielectric constant $\epsilon^{'}$ as measured at different magnetic fields, with sharp downturns being observed at the field induced transition. This is also seen as a peak in the magnetic field dependent pyroelectric current measurements as is shown in the main panel. Polarization as obtained after integrating the pyrocurrent as measured in different electromagnetic poling conditions is shown in (c).}       
	\label{Fig6}
\end{figure}
Our low temperature x-ray diffraction data also indicates that the onset of ferroelectric order is accompanied by an abrupt change in the in-plane Fe1-O2 and the Fe2-O1 bond lengths, with the former decreasing (and the later increasing) across this transition, as is shown in Fig.\ref{Fig4}(a). This represents a distortion of the polyhedral cage along the $ab$ plane in the fashion shown in Fig.\ref{Fig4}(b). Interestingly, this distortion does not appear to change the crystallographic structure, in the sense that the diffraction patterns on either side of the phase transition could be fit using the same Trigonal ($P\bar{3}c1$) symmetry.  With no apparent change in the overall crystallographic symmetry, it is interesting to note that instead of exhibiting linear magnetoelectricity like the other members of the $A_{4}$Ta$_{2}$O$_{9}$ family,  only this system appears to exhibits true polar order. We speculate that this could  arise from an effective reduction in the dimensionality of the magnetic lattice in this system. Our magnetization data clearly indicates that approximately half of the Fe${^{2+}}$ ions in Fe${_4}$Ta${_2}$O$_{9}$ stabilize in the nonmagnetic ($S$=0) state. In this context, it is interesting to note that the Fe${^{2+}}$ ions have two distinct crystallographic sites Fe1 ($1/3, 1/3, 0.517$) and Fe2 ($1/3, 2/3, 0.307$), which lie in different planes stacked along the crystallographic $c$ axis. As is shown in Fig. \ref{Fig5}, these two distinct lattice sites form separate distorted hexagonal sub-lattices along the $ab$ plane, with the Fe1 sub-lattice being slightly buckled in nature [Fig.\ref{Fig5}(a)] and the Fe2 sub-lattice having a zig-zag like character [Fig.\ref{Fig5}(b)]. If the different ($S$=2 and $S$=0) spin species of Fe${^{2+}}$ \emph{preferentially} occupy the two possible crystallographic sites, we would expect to see this being subtly reflected in the local octahedral environment of each of these FeO${_6}$ sub-units. Within an octahedral environment, a Fe${^{2+}}$ ($3d{^6}$) ion would be expected to have a $t{_{2g}}{^4}e{_g}{^2}$ and $t{_{2g}}{^6}e{_g}{^0}$ electronic configurations in the high spin and low spin states respectively. With the $d$ orbitals of the occupied $e{_g}$ electrons being along the direction of the oxygen $2p$ orbitals, the resultant Coulomb interaction would be expected to result in a slight elongation of the resultant bond, whereas the $t{_{2g}}$ orbitals being directed along the bisector of the O-Fe-O angle would be expected to have a smaller effect. Analysis of our x-ray diffraction data indicate that the average Fe1-O bond length ($2.16\AA$) exceeds that of the Fe2-O bond length ($2.05\AA$) by $5\%$. This leads us to believe that the magnetic ($S=2$) Fe${^{2+}}$ ions exclusively occupies the  buckled honeycomb sub-lattice, with the nonmagnetic ($S$=0) Fe${^{2+}}$ ions occupying the alternating zig-zag one.  This basically reduces the dimensionality of the magnetic lattice from a 3D to a quasi-2D one. It is well known that short range magnetic fluctuations can persist in a wide temperature window above the magnetic transition in 2D magnets \cite{ShortCorrel}, signatures of which we have also observed in our magnetization  measurements.  

Measurements of the dc magnetic susceptibility measured in magnetic fields in excess of 1 Tesla reveals the presence of an additional magnetic field induced transition. Appearing at $\approx$ 22 K at fields of the order of 1 Tesla, this temperature at which this feature appears increases as a function of the applied magnetic field, and reaches a maximum of $\approx$ 35 K, at the highest applied magnetic field of 7 Tesla. as is clearly evident from d($1/\chi$) /dT vs T plot in Fig. \ref{Fig6} (a). This field induced transition is also clearly discernible in the dielectric measurements performed in the presence of magnetic fields, in the form of a sharp downturn in the real part of dielectric constant  $\epsilon^{'}(T,H)$ as is seen in the inset of Fig.\ref{Fig6} (b). This magnetic field induced transition was further evaluated by means of pyroelectric current measurements performed  in the presence of different magnetic fields after typical magnetoelectric (ME)  poling. This  involved poling the specimen at different values of the electric ($E$) and magnetic ($H$) fields from a poling temperature of  50K, with $E \perp H$.  At low temperatures the electric field alone was removed, and standard pyroelectric measurements were performed in the presence of the magnetic field. As is shown in the main panel of Fig. \ref{Fig6} (b), this field induced transition also manifests itself in the form of a sharp peak in the pyrocurrent. The polarization associated with this magnetic field induced transition can be fully flipped by reversing the direction of the electric and magnetic fields, as depicted in Fig. \ref{Fig6} (c), for the case of measurements done at  $H$ = 7 Tesla, indicating the true ferroelectric nature of this field induced transition. 
  \newline 
  \begin{figure}
	\centering
	\includegraphics[scale=0.43]{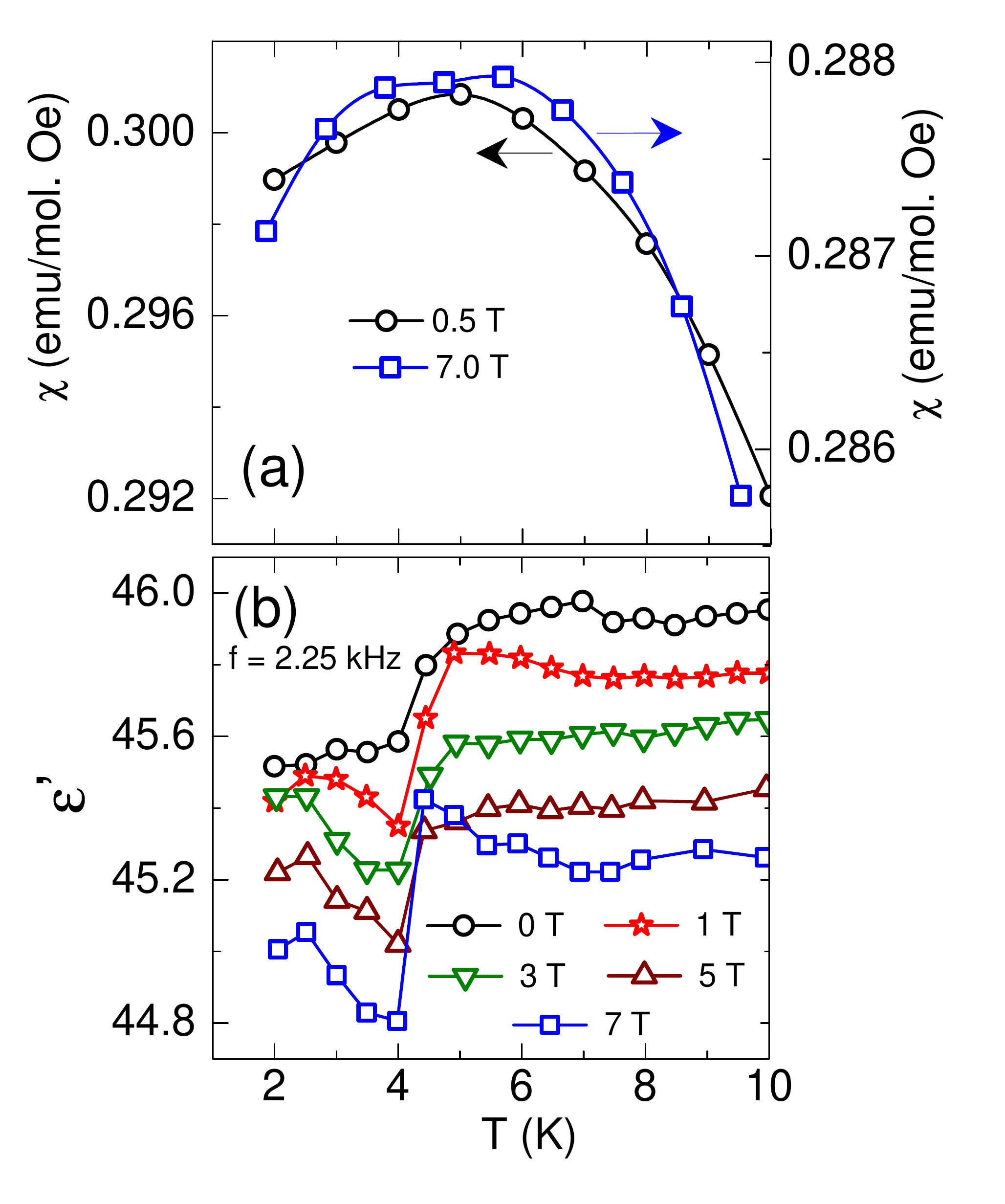}
	\caption{(color online) (a) The DC magnetic susceptibility as measured at 5 kOe, and 7 Tesla fields, indicating a cusp in the magnetization corresponding to a possible transition at 5K. (b) depicts measurements of the dielectric constant in the same temperature range, as measured at different magnetic fields. }
	\label{Fig7}
\end{figure}

Below this field induced magnetic state, the ZFC magnetic susceptibility increase monotonically until the magnetic anomaly at 5 K Fig. \ref{Fig7}(a). Though this magnetic transition is not discernible in our specific heat measurements, implying that the change in entropy associated with it is rather small, the transition is observed to be robust in the presence of large magnetic fields of the order of 7 Tesla.. Dielectric measurements (Fig.\ref{Fig7} (b)) indicate a step-like feature at this temperature, reminiscent of that seen in many systems across their magnetic transitions. Measurements done at different magnetic fields show that there is no observable change in the nature of the magneto-dielectricity on either side of this phase transition, suggesting that the nature of the polar state remains unchanged. This is also in agreement with our pyroelectric measurements which do not exhibit a peak in the pyrocurrent when the specimen is poled from a poling temperature of 15K, thus ruling the possibility of a new ferroelectric phase. Interestingly, on poling from temperatures above 35 K in the presence of large magnetic fields we observe that the polarization of the field induced state sustains well below 5K. This clearly indicates that the field induced ferroelectric state is not affected by this magnetic transition, and also reinforces our observation that this low temperature transition is not associated with a new polar state.  

\begin{figure}
	\centering
	\includegraphics[scale=0.43]{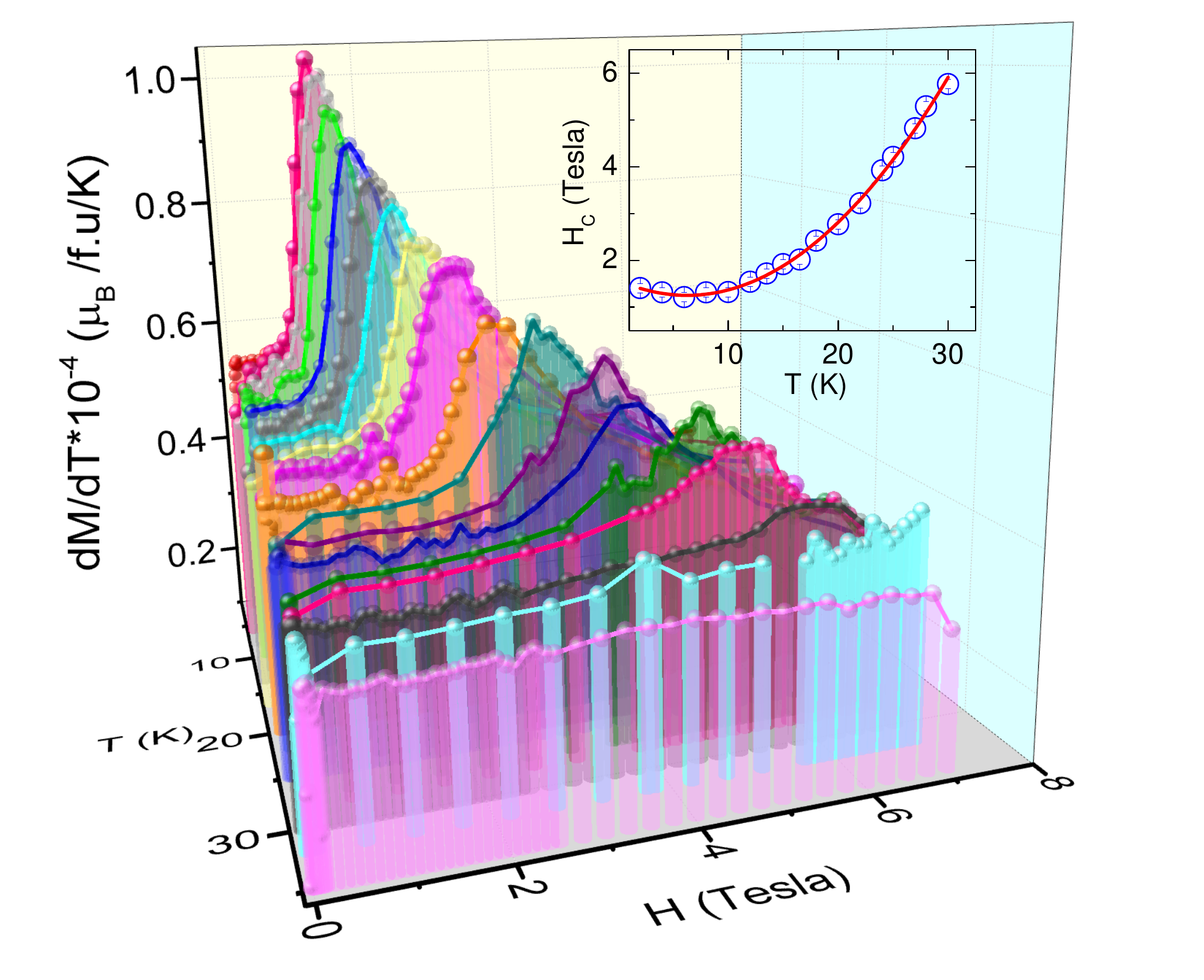}
	\caption{(color online)  $dM/dH$ of the $MH$ isotherms as measured at different temperatures. The inset depicts the temperature evolution of the critical field ($H_C$) of the metamagnetic transition, with the solid line being a fit of the form $AT{^2}$ + $BT$ + $C$.}
	\label{Fig8}
\end{figure}

Our low temperature $M$-$H$ isotherms exhibit a non-saturating behavior up to the highest applied field of 7 Tesla as is expected for antiferromagnetic systems. A finite opening of the hysteresis loop is seen, possibly due to the presence of a finite ferromagnetic contribution, as was also indicated by a low value of the Curie-Weiss temperature ($\theta{_{CW}}$). Interestingly, we also observe the presence of a field induced metamagnetic transition as evidenced by a change of slope in the $M$-$H$ isotherms, which is seen more clearly in plots of $dM/dT$ vs $T$ plots in the form of a peak (Fig. \ref{Fig8}). The critical field associated with this metamagnetic transition is 1.4 T ($\pm$0.1 T) at 2K, and steadily increases as a function of temperature. Interestingly, there is no apparent signature in the dielectric or pyroelectric measurements at the $H$-$T$ values corresponding to this metamagnetic temperature, indicating that the polar state remains relatively unaltered.  The critical field ($H_{C}$) associated with this transition has a slope $dH{_C}/T > 0$, and also appears to have a $T^2$ dependence, as is shown in the inset of Fig. \ref{Fig8}. This is in broad agreement with Yamada's theory for itinerant metamagnets \cite{Yamada}, which has been observed in a number of systems. However, it has to be borne in mind that this theory was developed for field induced paramagnetic-ferromagnetic phase transitions in itinerant magnets - and hence its applicability within the antiferromagnetically ordered state of a system like Fe${_4}$Ta${_2}$O$_{9}$ is suspect.
 
\begin{figure}
  	\centering
  	\includegraphics[scale=0.35]{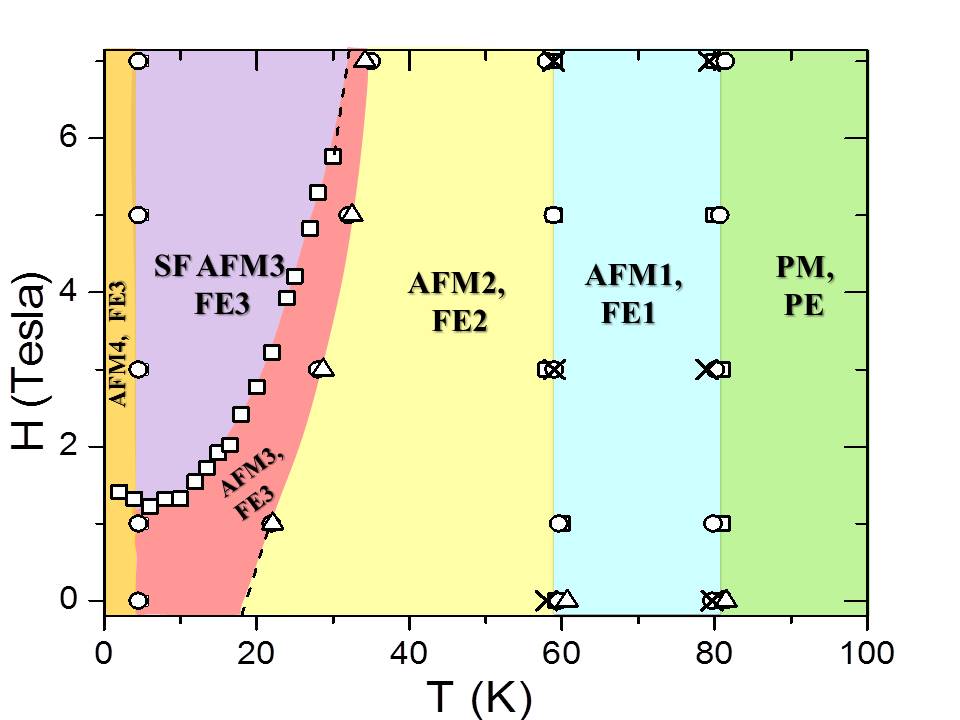}
  	\caption{(color online) The $H$-$T$ phase diagram of Fe$_4$Ta$_2$O$_9$ as determined from magnetic ($\square$), specific heat ($\times$), dielectric ($\circ$) and pyroelectric ($\bigtriangleup$) investigations. The dotted lines do not represent experimentally determined phase boundaries, and are only meant as a guide to the eye. }
  	\label{Fig9}
\end{figure}

The $H$-$T$ phase diagram of Fe${_4}$Ta${_2}$O$_{9}$ as determined using the magnetic, thermodynamic, dielectric and pyroelectric data at our disposal is depicted in Fig. \ref{Fig9}. As a function of decreasing temperature, this system undergoes a transition into two distinct multiferoic states labeled as AFM1 FE1, and AFM2 FE2 at 80K and 60K respectively. On cooling further, and in the presence of magnetic fields in excess of 1 Tesla, we observe a field induced multiferroic state denoted as AFM3 FE3. Within this state, we also observe the presence of a metamagnetic (spin-flop) transition which appears to have no discernible influence on the polar state (SF-AFM3 FE3).  At temperatures of the order of 5K, we further observe the onset of possibly a different kind of antiferromagnetic order, (AFM4 FE3) as evidenced from magnetic and dielectric measurements. Needless to say, in-field neutron diffraction measurements would be imperative to further reconfirm these phase boundaries, and to identify the changes in the magnetic structure across these different phases. It is remarkable that Fe${_4}$Ta${_2}$O$_{9}$ exhibits such a complex phase diagram, which is in stark contrast to the other members of the extended $A_{4}$Ta$_{2}$O$_{9}$ family. We note that the phase diagram of this system has some parallels to that observed in the Ni$_{3}$V$_{2}$O$_{8}$ \cite{NiVOJPCM, NiVOPhaseDia} and KCu${_3}$As${_2}$O${_7}$(OD)${_3}$ \cite{KCu} systems, stabilizing in distorted variants of the Kagome structure, where a series of low temperature magnetic phases coupled with polar order have been observed.  A common factor here appears to be the presence of 2 dimensional honeycomb layers which are buckled in and out of plane, with competing nearest and next-nearest neighbor interactions giving rise to a complex magnetic and polar phase diagram. 

\section{Conclusions}
In summary, using a combination of magnetic, thermodynamic, dielectric and pyroelectric measurements, we demonstrate that Fe${_4}$Ta${_2}$O$_{9}$ is a new multiferroic, in contrast to all the other members of its corundum related family of materials. We speculate that this arises as a consequence of the inherently two dimensional nature of the magnetically active sublattice that stabilizes in a buckled honeycomb structure. This system is also seen to exhibit a rich magnetic and polar phase diagram, and could hence provide a useful playground for investigating the complex interplay between nearest and next-nearest magnetic interactions, anisotropy and the Dzyaloshinskii-Moriya interactions in quasi 2 dimensional honeycomb lattices. 
\section{Acknowledgements}
The authors thank D. Buddhikot for his help in heat capacity measurements. J.K. acknowledges DST India for support through PDF/2016/000911. NPDF. S.N. acknowledges DST India for support through grant no. SB/S2/CMP-048/2013.
\bibliography{Bibliography}
\end{document}